\documentclass[%
 reprint,
 amsmath,amssymb,
 aps
floatfix,
]{revtex4-1}

\usepackage{color}
\usepackage{graphicx}
\usepackage{dcolumn}
\usepackage{bm}
\usepackage{subfigure}
\usepackage{caption}
\usepackage{natbib}
\usepackage[bookmarks=false]{hyperref}
{
\begin{document}

\preprint{APS/123-QED}

\title{On the Origin of Precipitation of Transition Metals Implanted in MgO
}

\author{Debolina Misra}
 \email{mm18ipf03@iitm.ac.in}
 \author{Satyesh K. Yadav}
\affiliation{
Department of Metallurgical and Materials Engineering, Indian Institute of Technology Madras, Chennai, 600036, India}

\date{\today}

\begin{abstract}
Transition metals implanted in single crystal MgO can precipitate out at grain boundaries or remain embedded in bulk. Using first-principles calculations based on density functional theory we have calculated the thermodynamic stability and diffusion coefficients of the implanted ions to explain Fe and Ni precipitation in MgO. Experimentally it has been observed that some of the Fe atoms precipitate out, while few Fe atoms in 2+ and 3+ charge states remain embedded in the lattice. Our simulation shows that at 600 K (typical annealing temperature) while neutral iron in MgO would migrate 1 $\mu$m in few microseconds, it takes several years for the charged Fe ions to migrate the same distance. On the other hand, Ni ions in all its charge states (neutral, 1+, 2+, and 3+) would migrate 1 $\mu$m in just few microseconds, at 600 K. This explains the experimental observation that implanted Ni always precipitates out. Our study paves a way forward to predict if ions implanted in stable oxide will be stable or will precipitate out. 
\end{abstract}

\maketitle


\section{\label{sec:intro}Introduction}
Implantation of metallic ions in refractory oxides have attracted immense attention \cite{hayashi_pssa} as the oxides find applications in switching and memory devices, spintronics and as dilute magnetic semiconductors \cite{MAO2007329}. Knowledge of the preferred charge states of implanted metals in the bulk oxides and their stability against precipitation at the grain boundaries are crucial for using the oxides in various devices. 
Stability and charge states of implanted metal ions in a range of oxides have earlier been studied experimentally with an aim to achieve specific applications \cite{WHITE1989, blas_prl, RAMOS1991, hao, elliman_2002, au_sio2, perez_1983, ZHU2006}. However, no attempts have been made to relate the thermodynamic stability of the implanted ions and their diffusivity in bulk oxides to the precipitation of the dopants at grain boundaries or their preferred charge states in the host oxide.
\\
This work relates the thermodynamic stability and diffusivity of Fe and Ni, implanted in MgO, to their observed charge states in the host, using first-principles method. Several studies on implanted Fe in MgO have revealed that some of the Fe atoms precipitate out while the rest Fe can remain embedded in the host lattice in 3+ and 2+ charge states\cite{WHITE1989, Molholt_2014, hayashi_pssa, MAO2007329}. For Ni on the other hand, available experiments show that Ni atoms implanted in MgO at room temperature get distributed in the matrix evenly; however, upon annealing, Ni precipitates out \cite{CRUZ2004840} with an average particle size of 8–10 nm \cite{ZHU2006}.
\\
It is assumed that Fe implanted in MgO substitutes lattice Mg atom and hence can be in 3+ and 2+ charge states \cite{Molholt_2014}. As substitutionally doped atoms have lower mobility and hence, more stability \cite{Wuensch}, Fe, if substitutes a lattice Mg atom can never precipitate out of MgO on annealing. In order to assess if implanted TM atom will substitute a lattice cation or will remain at interstitial site, in one of our previous work we showed that if ionic radius of TM is smaller than the host oxide cation, TM would be stable in the interstitial site \cite{sci_rep}; Fe and Ni came out to be stable as interstitials. Thermodynamic stability of Fe and Ni in MgO in various charge states was also calculated and we found that Fe can be stable in neutral, 2+ and 3+ charge states, while Ni can be stable in neutral 1+, 2+ and 3+ states as interstitial. 
In this work, we establish that Fe in neutral charge state and Ni in all charge states can easily diffuse through MgO and hence will easily precipitate out of MgO bulk. We also calculate diffusivity of Fe and Ni interstitials in various charges states in MgO, using first-principles density functional theory (DFT). 

\section{\label{sec:method}Methodology}
A cubic super cell containing 32 formula units of MgO with a dopant concentration of 3.1\% has been considered for studying the thermodynamic stability and diffusion of Fe and Ni dopants in the oxide. Density functional theory as implemented in Vienna Ab initio Simulation Package (VASP) \cite{kresse1996commat, kresse1996efficient} is used for all our calculations with projector-augmented wave (PAW) method \cite{blochl1994projector}. Generalized gradient approximation (GGA) was used to treat the exchange correlation interaction with the Perdew, Burke, and Ernzerhof (PBE) functional \cite{perdew1996generalized}. All the structures were fully relaxed using the conjugate gradient scheme and relaxations were considered converged when force on each atom was smaller than 0.02 eV/\AA. A plane wave cut-off of 500 eV and a k-point mesh of 5x5x5 were used for achieving converged results within 10$^{-4}$ eV per atom. The density of states (DOS) for the doped systems were calculated by the linear tetrahedron method with Bl\"{o}chl corrections and using a denser k-grid.

\section{Results}
Here we first briefly summarise the results on preferred site and charge states of Fe and Ni in MgO (which we had already reported earlier in detail) for the sake of completeness. Then we report our simulations on diffusion barrier of Fe and Ni in various charge states in MgO. 
\subsection{Thermodynamic Stability of Fe and Ni in MgO}
\begin{figure}[h!]
\centering
\includegraphics[height=2.4in]{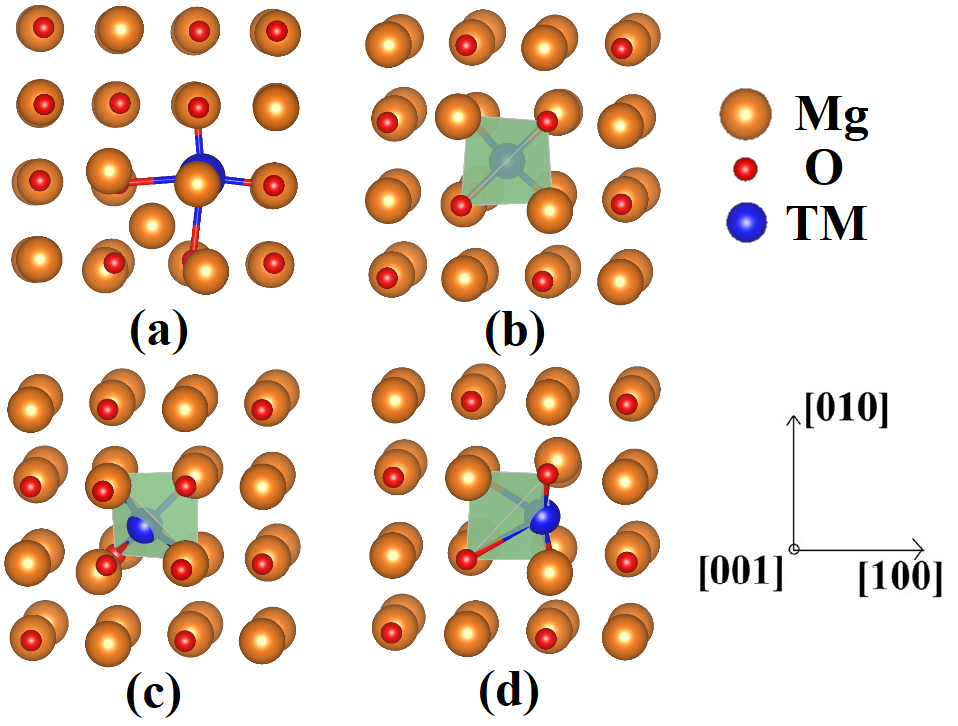}
\caption{Atomic structure of TM dopants (Fe or Ni) in MgO: (a) TM replaces host cation and the host cation sits in the tetrahedral interstitial; TM dopant sits at the (b) middle (c) corner and (d) edge of the oxygen tetrahedra.} 
{\label{structure}}
\end{figure}
To explore whether Fe and Ni will occupy interstitial site or substitutional site, energies of the two configurations were compared: Fe or Ni (i) substituting lattice Mg atom (wyckoff 4a(0.5, 0, 0.5)) and pushing it to the center of the nearest oxygen tetrahedra (Fig.\ref{structure}(a)) and (ii) occupying tetrahedral interstitial site (wyckoff 8c(0.25, 0.25, 0.25)) (Fig.\ref{structure}(b)) \cite{sci_rep}. If the first configuration comes out to be more stable, Mg atom would be substituted by Fe or Ni. Fe or Ni can be stable in the host as an added atom if the second configuration is more stable. While calculating the substitutional formation energy, the substituted lattice cation is generally removed from the host to maintain overall stoichiometry. However, this approach holds valid when doping is achieved via conventional chemical synthesis routes. For doping through ion implantation, there can be metal added to the host lattice if the dopant occupies the interstitial site. 

\begin{figure}
\centering
\includegraphics[height=3.6in]{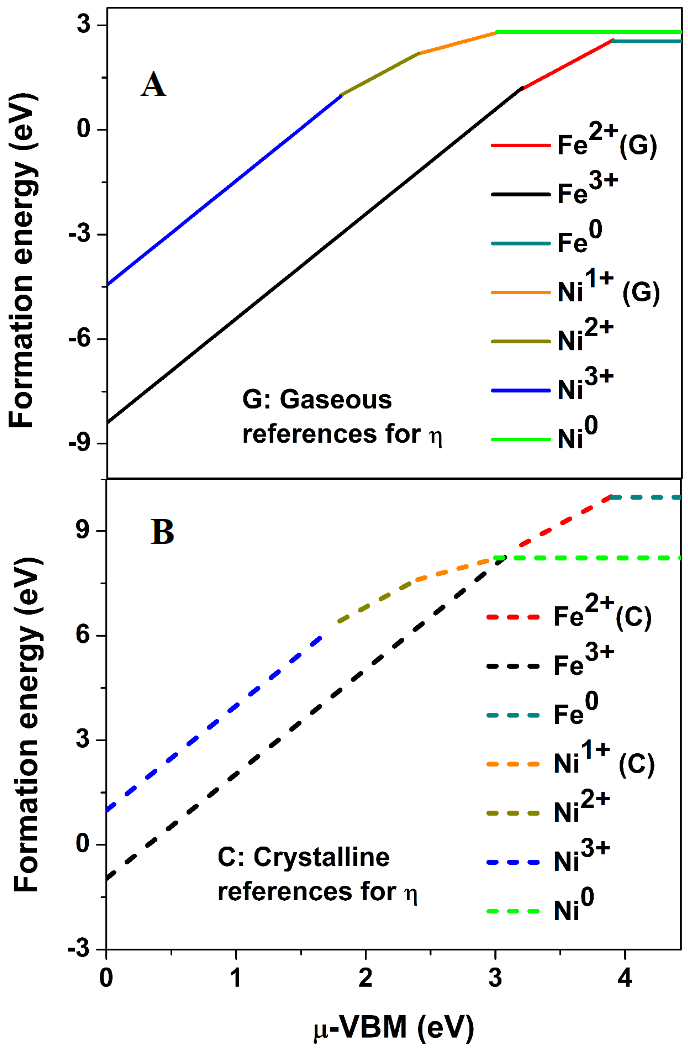}
\caption{Formation energy of the neutral and charged Fe and Ni atoms in MgO as a function of electronic chemical potential $\mu$. Chemical potential ($\eta$) of Ni and Fe  (A) gaseous and (B) crystalline.} 
{\label{charge}}
\end{figure}
The charge states preferred by Fe and Ni while occupying the interstitial sites in MgO can be found in terms of the dopant formation energy $E_f^q$ \cite{Rampiprl, Freysoldtprb2016, VandeWalleRev2014, BAJAJ2015} 
\begin{equation}
E_f^q=E_D^q-E_B-\eta+q(\mu+E_{ref}+{\Delta}V)+E_{corr}^q
\end{equation}
Here \textit{$E_D^q$} and \textit{$E_B$} represent the total energies of defect supercell with charge \textit{q} and the defect free host supercell, respectively. \textit{$\eta$} is the chemical potential of the transition metal atom species. The '-' sign indicates addition of the TM defect in host. \textit{$E_{ref}$} is a suitable reference energy, taken to be the valence band maximum (VBM)\cite{sci_rep} of the oxide. \textit{$\mu$} is the electronic chemical potential of the system that varies from VBM up to the band-gap of MgO obtained from our DFT calculation. \textit{$\Delta$V} is the correction necessary to realign the reference potential of the defect supercell with that of the defect free supercell \cite{VandeWallejap2004} and $E_{corr}^q$ is the first-order monopole correction to the electrostatic interaction and the finite size of the supercell.
\\
We showed that both Fe and Ni prefer to occupy tetrahedral interstitial sites in MgO irrespective of their charge states. However, neutral Ni occupies not the center but one corner of the oxygen tetrahedron (wyckoff 32f (0.81, 0.688, 0.688))(Fig.\ref{structure}c). Neutral Fe on the other hand prefers to sit in between two oxygen atoms forming the tetrahedron (wyckoff 48 g (0.25, 0.9, 0.25)) (Fig.\ref{structure}d). Fe ion occupying the edge of the oxygen tetrahedra in MgO is also supported by an earlier M\"{o}ssbauer spectroscopic analysis of Fe in MgO \cite{Molholt_2014}.
\\
Figure \ref{charge} shows formation energy of Fe and Ni in various charge states as a function of electronic chemical potential. Both the gaseous and crystalline metal energy references for the chemical potential of Fe and Ni are taken into account, where gaseous reference to chemical potential indicates Fe and Ni energies in atomic state. While formation energies obtained with gaseous reference is necessary to access the thermodynamic driving force of TM atoms when implanted, crystalline energy of Fe and Ni as reference is more relevant when chance of precipitation of the implanted ions is to be investigated. Our calculations shows that only Fe$^{3+}$ has negative formation energy which indicates that Fe should be observed only in 3+ charge state. Besides this, our calculated formation energies also suggest that Fe in neutral and 2+ charge states, and Ni in all charge states should precipitate out. Hence in order to have a thorough understanding of the implanted Ni and Fe in MgO, we further proceeded to probe the diffusion behaviour of these ions in the host lattice. 
\begin{table}[h!]
\setlength{\tabcolsep}{4pt}
  \begin{center}
  \caption{Attempt frequencies \textit{$\nu$} and transition barriers \textit{E$_{ij}$} (from \textit{i}th to \textit{j}th interstitial) for Fe and Ni in MgO. Debye temperature \textit{T$_D^{MgO}$}=750 K and Debye frequency \textit{$\nu_D^{MgO}$} = 15.62 THz. \textit{t$_D$} is the time required for ions to diffuse 1 micrometer at \textit{T}=600 K.}
    \label{tab:barrier}
     \begin{tabular}{c|c|c|c} 
      \multicolumn{1}{c|}{Attempt frequencies ($\nu$)} & \multicolumn{1}{c|}{Dopant} & \multicolumn{1}{c|}{$E_{ij}$(eV)} & \multicolumn{1}{c}{time(\textit{t$_D$})}\\
      \hline
      & Fe$^0$ & 0.15 & 7.72x10$^{-8}$ Sec\\
     $\nu_{Fe}$ = 13.28 THz& Fe$^{2+}$ & 2.16 & 94 Years\\
      & Fe$^{3+}$ & 2.21 & 246 Years\\
      \hline
      & Ni$^0$ & 0.02 & 3.2x10$^{-9}$ Sec\\
      $\nu_{Ni}$ = 12.95 THz& Ni$^{1+}$ & 0.08 & 1.02x10$^{-8}$ Sec\\
      & Ni$^{2+}$ & 0.11 & 1.83x10$^{-8}$ Sec\\
      & Ni$^{3+}$ & 0.59 & 1.96x10$^{-4}$ Sec\\
   \end{tabular}
  \end{center}
\end{table}
\begin{figure}[h!]
\centering
\includegraphics[height=2in]{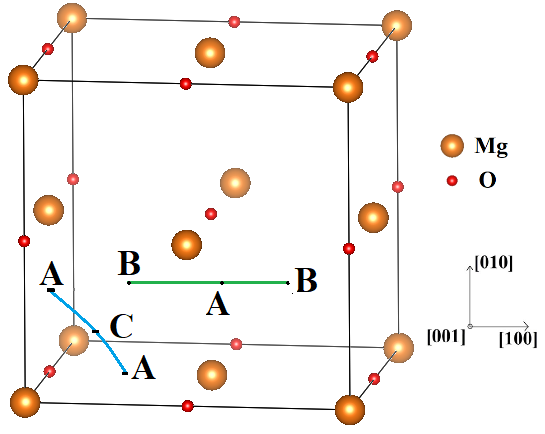}
\caption{Diffusion paths for Fe and Ni dopants in MgO unit cell. Fe$^{2+}$, Fe$^{3+}$, and Ni in all charge states follow B--A--B path for diffusion. Fe$^0$ has two possible paths for diffusion:  A to another equivalent A site (i) via C or (ii) through B. The barriers for transition for Fe$^0$ via the paths A--C--A and A--B--A came out to be 0.15 eV and 1.29 eV respectively, indicating that A--C--A path is preferred for diffusion. } 
{\label{fe0}}
\end{figure}
\subsection{Barrier of transition and diffusion coefficient}
Hopping of Fe and Ni from one lattice site to the other leads to their precipitation at the grain boundary. DFT provides a reliable way of calculating this transition barrier which is essential for estimating the time TM dopants take to diffuse out of MgO. Using Climbing Image Nudged Elastic Band (CI-NEB) method \cite{ci-neb}, barrier of transition between nearest interstitial sites have been calculated for both Fe and Ni in various charge states as listed in Table\ref{tab:barrier}.
Two possible diffusion paths are shown in Fig.\ref{fe0}; while Fe$^0$ occupies the A site, rest of ions occupy B site. Among the two possible pathways for diffusion of neutral Fe, we find that it is more likely for Fe$^0$ to follow the path A--C--A as the barrier height of transition is substantially low (0.15 eV) compared to A--B--A path with a sufficiently high (1.29 eV) transition barrier. However, for Fe$^{3+}$ and Fe$^{2+}$, barrier heights for transition are more than 2 eV. Ni, on the other hand, has very low barriers for diffusion in all charge states, with a maximum barrier height being 0.59 eV for Ni$^{3+}$. In general, a heigh barrier of transition is indicative of the fact that the ion under consideration is stable in the host lattice.  Activation energies of Fe$^{3+}$ and Fe$^{2+}$ in interstitial site are comparable to the experimentally reported activation energies for substitutionally doped TM dopants (1.81 eV for Fe$^{3+}$ and 2.10 eV for Ni$^{2+}$)\cite{Wuensch}. However, the activation energies of Fe$^0$ and Ni in all charge state are significantly lower. 
For calculating the diffusion coefficients we have adopted a method proposed by Wu et al \cite{wu_diff}. Transition frequencies $\lambda_{ij}$ are first computed from
\begin{equation}
\lambda_{ij} = \nu_{ij} exp(-E_{ij}/k_BT)
\end{equation}
Here \textit{$\nu_{ij}$} is the attempt frequency and \textit{E$_{ij}$} is the barrier of transition between sites \textit{i} and \textit{j}. \textit{k$_B$} and \textit{T} are Boltzmann constant and temperature respectively. For diffusion of Fe and Ni, \textit{$\nu_{ij}$} can be approximated using the relation
\begin{equation}
\nu_{ij} = \nu_D \sqrt{m_{matrix}/m_{TM}}
\end{equation}
where \textit{$\nu_D$} is the Debye frequency of the host which can be easily calculated from its Debye temperature (\textit{$T_D = \nu_Dh/k_B$}). \textit{m$_{matrix}$} and \textit{m$_{TM}$} refer to the masses of MgO and TM (Fe, Ni) dopants respectively, and \textit{h} is the Planck's constant. Attempt frequency values for Fe and Ni in MgO are listed in Table\ref{tab:barrier}.
\\Considering equal probabilities for diffusion along all three directions, the diffusion coefficients (\textit{D}) can be calculated from
\begin{equation}
D = \frac{1}{6}\lambda_{ij}\alpha^2
\end{equation}
where \textit{$\alpha$} is the length of diffusion. For our calculations, alpha is length of A--B--A path or A--C--A path. 
\begin{figure}[h!]
\centering
\includegraphics[height=4in]{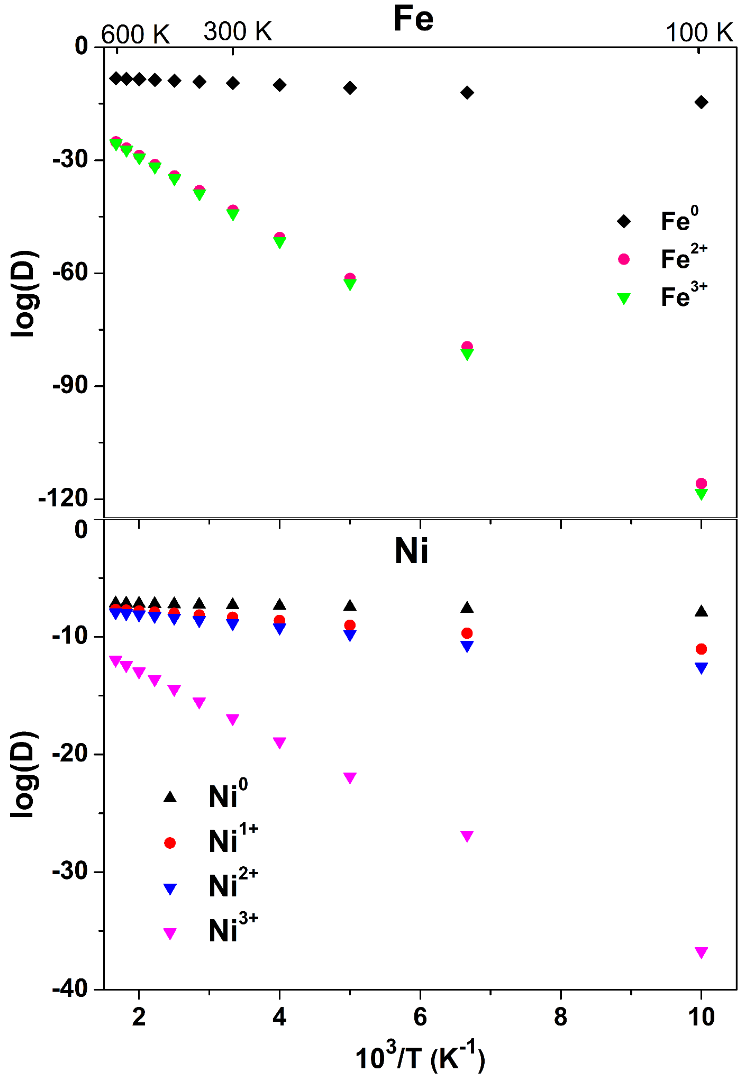}
\caption{Diffusivity of charged Fe and Ni dopants in MgO as a function of temperature.} 
{\label{diff_co}}
\end{figure}

Our DFT calculated diffusion coefficients for neutral and charged Fe and Ni in MgO are shown in Figure\ref{diff_co}. Fe$^{3+}$ has the lowest diffusivity among all and is comparable to that of Fe$^{2+}$. This suggest that both Fe$^{3+}$ and Fe$^{2+}$ share similar diffusion behaviour. On the other hand, diffusivity of Ni in all charge state and Fe in neutral state are several orders higher than Fe$^{3+}$ and Fe$^{2+}$, even at 600 K which suggests that a much slower diffusion of Fe in 2+ and 3+ charge states should be expected than Fe$^0$ and Ni ions. 
\\
We also estimate the time required for ions to diffuse to 1 micrometer (typical grain size) at 600 K (typical annealing temperature) from the speed of diffusion which can be calculated as
\begin{equation}
S=\frac{1}{6}\alpha\lambda_{ij}
\end{equation} 
assuming that it takes 6/$\lambda_{ij}$ seconds for the ions to traverse the transition path length of $\alpha$. The time required to diffuse 1 micrometer by various ions are listed in Table\ref{tab:barrier}. While Ni in all charge states and Fe in neutral state can diffuse 1 micrometer in less than a second, it takes several years for Fe in 2+ and 3+ charge states to diffuse the same length. Upon annealing, Ni (in all charge states) and neutral Fe being unstable in MgO, precipitate out of MgO very quickly. However, Fe$^{2+}$ despite being unstable can remain in MgO, as it takes several years for Fe$^{2+}$ to diffuse out to the grain boundary. 
\begin{figure} [h!]
\centering
\includegraphics[height=2.9in]{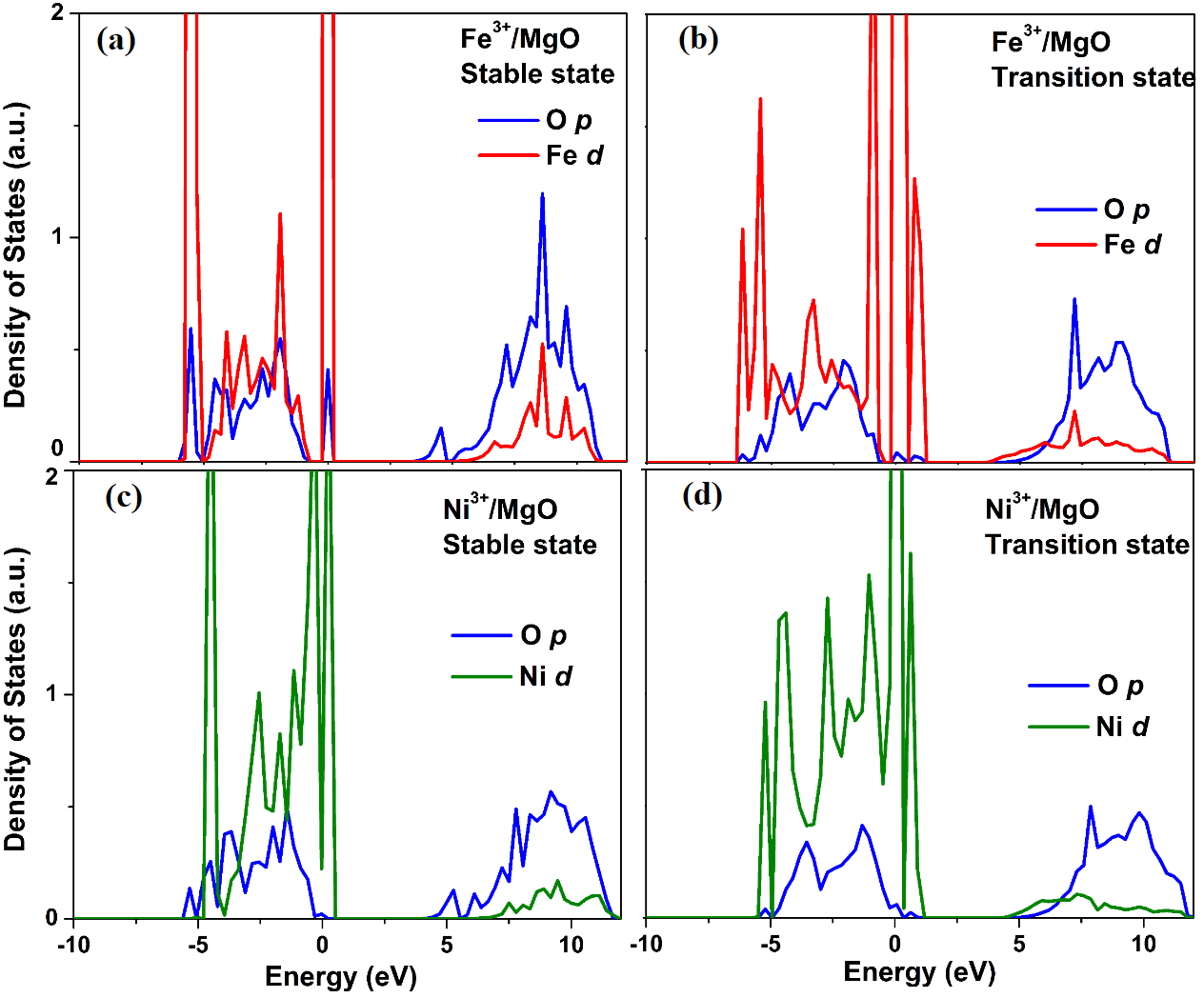}
\caption{Density of states plots for Fe$^{3+}$ and Ni$^{3+}$ in MgO in initial and transition states. \textit{d} states for Fe and Ni and \textit{p} states for neighboring oxygen atoms are shown.} 
{\label{dos}}
\end{figure}

\subsection{ Electronic structure }
To study if barrier of diffusion of the TM dopants in MgO also gets reflected from their respective electronic density of states (DOS), orbital resolved DOS for doped MgO have been calculated. Figure\ref{dos} shows the \textit{d}-states of Fe and Ni both in 3+ charge states and the \textit{p} states of neighbouring O-atoms. From Fig\ref{dos}(a)-(b) it is clearly evident that from the initial to transition state, the peaks shift to the higher energy region and the hybridization between \textit{p} and \textit{d} states also decreases, indicating the lower stability of the later configuration. The plot also reveals the changes in the bonding between Fe and Ni \textit{d} with O \textit{p} in the transition states. The decrease in TM \textit{d} and O \textit{p} hybridization from initial to transition state, is more in Fe than in Ni, which in turn explains higher diffusion barrier for Fe$^{3+}$. A comparison between Fe$^{3+}$ and Ni$^{3+}$, both in their respective initial positions, reveals that hybridization between Ni \textit{d} and O \textit{p} states is substantially low compared to the same between Fe$^{3+}$ and neighbouring O atoms. This clearly indicates that in terms of overall stability, Fe$^{3+}$ is much stable than Ni$^{3+}$, as already seen from our defect formation energy calculations.

\section{Conclusions}
This work explains experimentally observed precipitation of Fe and Ni using our DFT calculated thermodynamic stability and diffusivity of Fe and Ni interstitial in MgO. We also explained why Fe$^{2+}$ and Fe$^{3+}$ ions are observed in MgO when implanted with Fe. We showed that rather than substituting lattice Mg atoms, Ni and Fe prefer to be in the interstitial site of MgO in all charge states. Our calculations show that, Ni in all charge states, and Fe in neutral and 2+ charge states are unstable, while Fe$^{3+}$ is stable in MgO. What favours precipitation of Ni and neutral Fe is not only their instability but also their high diffusivity in MgO. At 600 K it takes Ni ions and Fe in neutral state less than a second to diffuse 1 micrometer. Fe$^{2+}$ on the other hand despite being unstable can remain in MgO as the required barrier for diffusion is high and comparable to substitutional Fe and it will take several years for Fe$^{2+}$ to diffuse 1 micrometer in MgO. Our finding opens a new way for predicting the preferred charge states of implanted metal ions in oxides and their stability against precipitation.  

\begin{acknowledgements}
 D.M. gratefully acknowledges the support from institute post-doctoral fellowship provided by IIT Madras, India. We acknowledge Dr. Somnath Bhattacharya's help for providing access to VASP source code.
\end{acknowledgements}

\bibliographystyle{ieeetr}
\bibliography{paper}

\end{document}